\documentclass[twocolumn,floatfix,showpacs,eqsecnum,prb,superscriptaddress]{revtex4}
\usepackage{graphicx}
\usepackage{amsmath}
\usepackage{amssymb}
\usepackage{bm}

\usepackage{color}

\begin{document}

\title{Thermal metal-insulator transition in a helical topological superconductor}
\author{I. C. Fulga}
\affiliation{Instituut-Lorentz, Universiteit Leiden, P.O. Box 9506, 2300 RA Leiden, The Netherlands}
\author{A. R. Akhmerov}
\affiliation{Instituut-Lorentz, Universiteit Leiden, P.O. Box 9506, 2300 RA Leiden, The Netherlands}
\author{J. Tworzyd{\l}o}
\affiliation{Institute of Theoretical Physics, Faculty of Physics, University of Warsaw, \\ Ho\.{z}a 69, 00--681 Warsaw, Poland}
\author{B. B\'{e}ri}
\affiliation{TCM Group, Cavendish Laboratory, J. J. Thomson Avenue, Cambridge CB3 0HE, United Kingdom}
\author{C. W. J. Beenakker}
\affiliation{Instituut-Lorentz, Universiteit Leiden, P.O. Box 9506, 2300 RA Leiden, The Netherlands}
\date{May 2012}
\begin{abstract}
Two-dimensional superconductors with time-reversal symmetry have a $\mathbb{Z}_{2}$ topological invariant, that distinguishes phases with and without helical Majorana edge states. We study the topological phase transition in a class-DIII network model, and show that it is associated with a metal-insulator transition for the thermal conductance of the helical superconductor. The localization length diverges at the transition with critical exponent $\nu\approx 2.0$, about twice the known value in a chiral superconductor.
\end{abstract}
\maketitle

\section{Introduction}
\label{intro}

Gapped electronic systems are topological states of matter, subject to phase transitions in which a topological invariant ${\cal Q}$ changes its value.\cite{Ryu10c} For noninteracting electrons in two spatial dimensions, ${\cal Q}\in\mathbb{Z}$ is integer valued in the absence of time-reversal symmetry, and ${\cal Q}\in\mathbb{Z}_{2}$ is binary in its presence. Two familiar examples from semiconductor physics are quantum Hall (QH) insulators in a strong magnetic field and quantum spin-Hall (QSH) insulators in zero magnetic field.\cite{Has10,Qi11} The topological invariant ${\cal Q}$ determines the number of electrically conducting edge states, which changes by $\pm 1$ at a topological phase transition. The edge states are chiral (unidirectional) in the QH effect and helical (counterpropagating) in the QSH effect.

Superconductors can also have an excitation gap, with topologically protected edge states in two dimensions. The edge states carry heat but no charge, so superconducting analogues of the quantum (spin)-Hall effect refer to the thermal rather than the electrical conductance. The thermal quantum Hall effect (${\cal Q}\in\mathbb{Z}$) appears in the absence of time-reversal symmetry, for example, in a single layer of a chiral \textit{p}-wave superconductor.\cite{Rea00,Sen00,Vis01,note1} Time-reversal symmetry is restored in two layers of opposite chirality $p_{x}\pm ip_{y}$, producing the thermal analogue of the quantum-spin Hall effect (${\cal Q}\in\mathbb{Z}_{2}$).\cite{Ryu10b,Wan11,Nom12,Sto12} The edge states in both effects are Majorana fermions, chiral for the thermal QH effect and helical for the thermal QSH effect.

There is a large literature on topological phase transitions in QH and QSH insulators, as well as in chiral superconductors,\cite{Eve08} but the thermal QSH effect in helical superconductors has remained largely unexplored. This is symmetry class DIII, characterized by the absence of spin-rotation symmetry and the presence of both time-reversal and electron-hole symmetry. Here we present a study of the phase diagram and critical behavior in helical superconductors. We use a network model in symmetry class DIII for a numerically efficient approach. 

We find that the main qualitative effect of time-reversal symmetry is that the transition between two topologically distinct thermal insulators goes via a thermal metal phase, for finite but arbitrarily weak disorder. In contrast, without time-reversal symmetry (in class D) the value of ${\cal Q}$ changes directly without an intermediate metallic phase for weak disorder. For strong disorder both chiral and helical superconductors have a thermal metal-insulator transition, but the critical behavior is different: We find a localization length exponent $\nu\approx 2.0$, about twice as large as the known value for chiral \textit{p}-wave superconductors.\cite{Med10}

The outline of this paper is as follows. To put our results for helical superconductors in the proper context, in the next section we first summarize known results for chiral \textit{p}-wave superconductors. In Sec.\ \ref{model} we introduce the network model of a helical superconductor, constructed out of two coupled chiral networks.\cite{Cho97} To identify topologically distinct phases we apply a scattering formulation of the topological quantum number,\cite{Ful12} as described in Sec.\ \ref{topquantnum}. In Sec.\ \ref{topphasetrans} we then present the results of our investigation: the phase diagram with the thermal metal-insulator transition, the scaling of the thermal conductivity at the transition, and the critical exponent for the diverging localization length. We conclude in Sec.\ \ref{conclude}.
 
\section{Chiral versus helical topological superconductors}
\label{DvsDIII}

According to the Altland-Zirnbauer classification,\cite{Alt97} superconductors
without spin-rotation symmetry are in class D or DIII depending on whether
time-reversal symmetry is broken or not. In two dimensions both symmetry
classes can be in thermally insulating phases which are topologically distinct
(with or without edge states). In this section we summarize what is known for
the phase diagram in class D, before turning to the effects of time-reversal
symmetry in class DIII.

A simple model Hamiltonian in class D represents a chiral \textit{p}-wave superconductor in the \textit{x-y} plane,
\begin{equation}
H_{\rm D}=v_{\Delta}(p_{x}\tau_{x}+p_{y}\tau_{y})+\left(\frac{p^{2}}{2m}+U-\mu\right)\tau_{z}.\label{HDdef}
\end{equation}
The Pauli matrices $\tau_{i}$ (with $\tau_{0}$ the $2\times 2$ unit matrix) operate on the electron-hole degree of freedom, coupled by the pair potential $v_{\Delta}(p_{x}\pm ip_{y})$. The Fermi energy is $\mu$ and $U(x,y)$ describes a random disorder potential (zero average). 

By adding a spin degree of freedom (with Pauli matrices $\sigma_{i}$ and unit matrix $\sigma_{0}$), one can extend
$H_{\rm D}$ to the Hamiltonian of a helical superconductor in class DIII,
\begin{align}
H_{\rm DIII}={}&v_{\Delta}(p_{x}\tau_{x}\sigma_{z}+p_{y}\tau_{y}\sigma_{0})+\left(\frac{p^{2}}{2m}+U-\mu\right)\tau_{z}\sigma_{0}\nonumber\\
&+K\tau_{y}\sigma_{y}.\label{HDIIIdef}
\end{align}
Both Hamiltonians have electron-hole symmetry, $\tau_{x}H^{\ast}\tau_{x}=-H$, but only the Hamiltonian \eqref{HDIIIdef} has time-reversal symmetry, $\sigma_{y}H^{\ast}_{\rm DIII}\sigma_{y}=H_{\rm DIII}$. The term $K\tau_{y}\sigma_{y}$ in $H_{\rm DIII}$ couples the two spin directions to zeroth order in momentum. Higher-order terms, such as $p\tau_{z}\sigma_{y}$, can be included as well.

\begin{figure}[tb] 
\centerline{\includegraphics[width=0.9\linewidth]{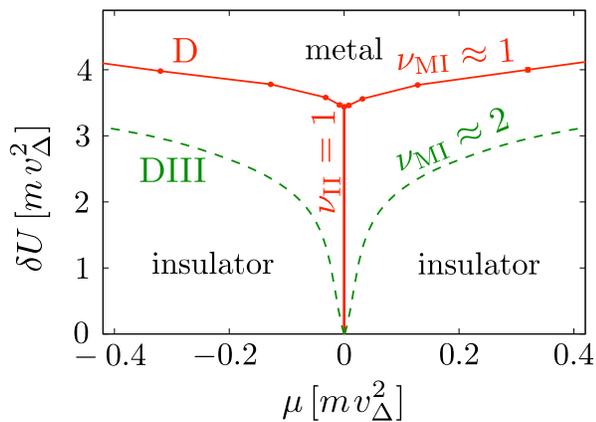}}
\caption{\label{fig_Dphase} Solid curves (red): Phase boundaries of a chiral \textit{p}-wave superconductor, symmetry class D, calculated in Ref.\ \onlinecite{Med10} from a lattice model based on the Hamiltonian \eqref{HDdef}.\cite{note2} The system is a thermal metal (Majorana metal) for strong disorder and a thermal insulator for weak disorder. The dashed (green) lines show qualitatively the effect of time-reversal symmetry in class DIII: the transition between two topologically distinct insulators then goes via an intervening metallic phase. The critical exponents for the various phase transitions are indicated: the value for $\nu_{\rm II}$ is exact,\cite{Lud94} the values for $\nu_{\rm MI}$ in class D (red, from Ref.\ \onlinecite{Med10}) and in class DIII (green, from this work) are numerical estimates.
}
\end{figure}

The phase diagram of the chiral \textit{p}-wave superconductor was calculated for a lattice model in Ref.\ \onlinecite{Med10}, by discretizing the Hamiltonian \eqref{HDdef} on a lattice. A similar phase diagram was obtained earlier\cite{Cha01,Mil07,Kag08} for a class-D network model (Cho-Fisher model).\cite{Cho97} As shown in Fig.\ \ref{fig_Dphase}, there are two insulating phases plus a metallic phase at strong disorder. The two insulating phases are topologically distinct, one is with and the other without chiral edge states. The disorder-induced thermal metal (Majorana metal) arises because of resonant transmission through zero modes (Majorana fermions), pinned to potential fluctuations where $U$ changes sign.\cite{Wim10,Kra11,Lau12}

The I-I phase boundary separating the two insulating phases and M-I phase boundary separating insulating and metallic phases meet at a tri-critical point. In the insulating phases the thermal conductivity decays exponentially $\propto e^{-L/\xi}$ with system size $L$. The localization length $\xi$ diverges $\propto |\mu-\mu_{c}|^{-\nu}$ on approaching a phase boundary at $\mu=\mu_{c}$. The critical exponent on the I-I phase boundary (at $\mu_{c}=0$) is known analytically,\cite{Eve08,Lud94} $\nu_{\rm II}=1$. The numerically obtained value\cite{Med10} $\nu_{\rm MI}=1.02\pm 0.06$ on the M-I phase boundary is very close to $\nu_{\rm II}$. Indeed, one would expect\cite{Lee11}  $\nu_{\rm MI}=\nu_{\rm II}$ if the phase boundaries at the tricritical point meet at nonzero angle, which they seem to do.

So much for a summary of known results for chiral superconductors in class D, without time-reversal symmetry. The time-reversally symmetric Hamiltonian $H_{\rm DIII}$ in \eqref{HDIIIdef} is just two uncoupled copies of $H_{\rm D}$ if $K=0$. Upon increasing the coupling strength $K$, the time-reversal symmetry starts to qualitatively modify the phase diagram. As we will show in what follows (as is indicated schematically in Fig.\ \ref{fig_Dphase}), a metallic phase develops in between the two insulating phases at weak disorder for $K\neq 0$.

\section{Class DIII network model}
\label{model}

For numerical efficiency we use a network representation of the class DIII Hamiltonian \eqref{HDIIIdef}. Network models\cite{Kra05} exist for quantum (spin)-Hall insulators\cite{Cha88,Obu07} and for chiral superconductors.\cite{Cho97,Kag99} A network model for helical superconductors was still missing and is provided here.

\subsection{Construction}
\label{construct}

\begin{figure}[tb] 
\centerline{\includegraphics[width=0.9\linewidth]{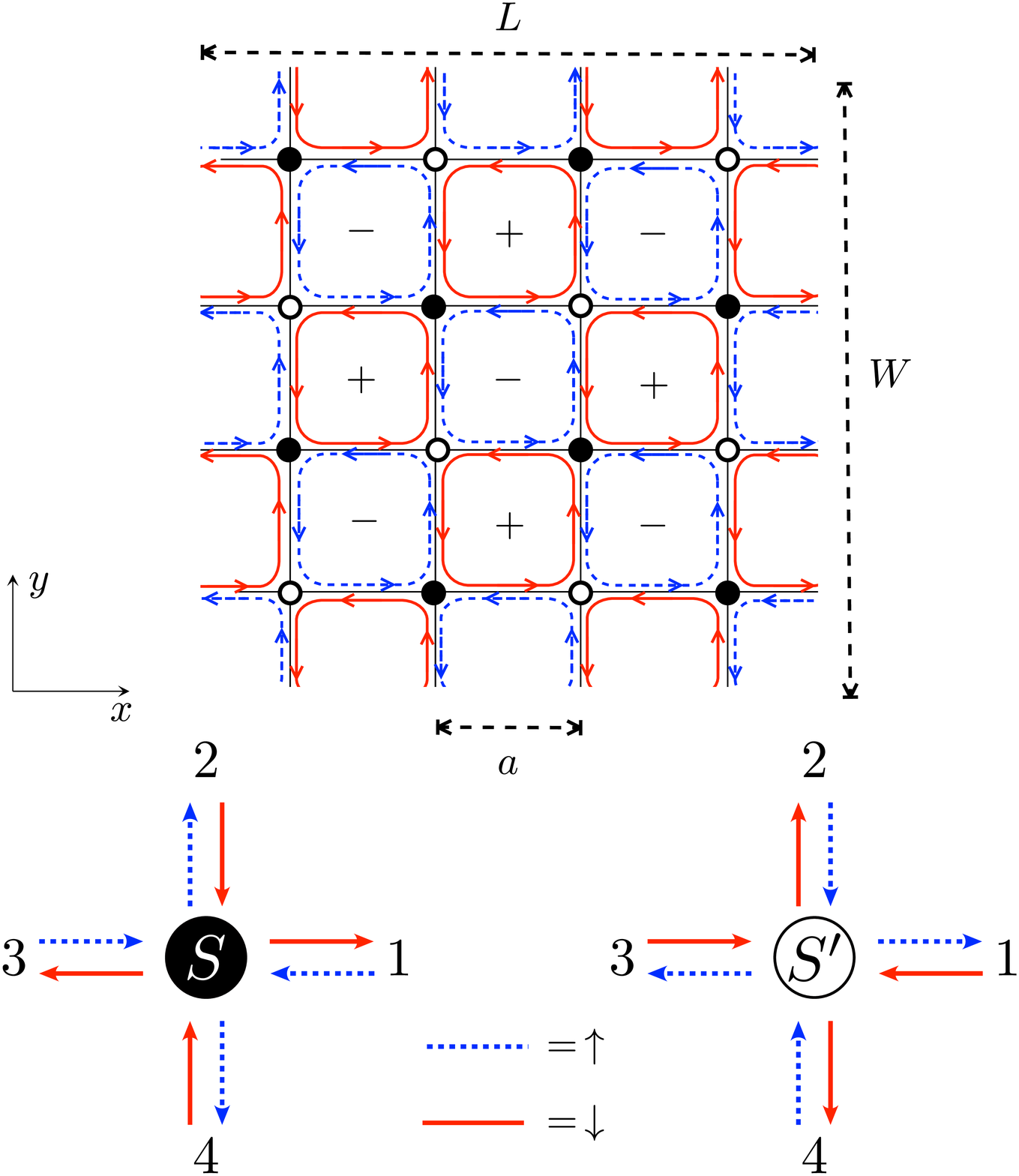}}
\caption{\label{fig:nodes} Illustration of the network model described in the text.
}
\end{figure}

The network is defined on a two-dimensional bipartite square lattice, see Fig.\ \ref{fig:nodes}. Helical Majorana modes propagate along the bonds and are scattered at the nodes. The helicity of the modes signifies that the direction of motion is tied to the spin degree of freedom, $\uparrow$ and $\downarrow$, represented by dashed and solid lines in the figure. The modes encircle local maxima and minima of the electrostatic potential (indicated by $\pm$ in the figure), in a clockwise or counterclockwise direction depending on the spin. The Majorana character of the modes signifies that there is no separate electron or hole mode, but one single charge-neutral mode per spin direction. 

The nodes of the lattice are saddle points between the local potential maxima and minima, alternating between adjacent plaquettes in a checkerboard pattern. Scattering at the nodes is described by $4\times 4$ unitary scattering matrices $S$ and $S'$ that alternate between adjacent nodes (black and white dots in the figure). The amplitudes $a_{n\sigma},b_{n\sigma}$ of incoming and outgoing modes are related by
\begin{equation}
 \begin{pmatrix}
  b_{1\downarrow} \\ b_{2\uparrow} \\ b_{4\uparrow} \\ b_{3\downarrow}
 \end{pmatrix} = S
 \begin{pmatrix}
  a_{1\uparrow} \\ a_{2\downarrow} \\ a_{4\downarrow} \\ a_{3\uparrow}
 \end{pmatrix},\;\;
\begin{pmatrix}
  b_{1\uparrow} \\ b_{2\downarrow} \\ b_{4\downarrow} \\ b_{3\uparrow}
 \end{pmatrix} = S'
 \begin{pmatrix}
  a_{1\downarrow} \\ a_{2\uparrow} \\ a_{4\uparrow} \\ a_{3\downarrow}
 \end{pmatrix}.\label{eq:spgrading}
\end{equation}
(The labels $\sigma=\uparrow,\downarrow$ and $n=1,2,3,4$ refer to Fig.\ \ref{fig:nodes}.)

A class DIII scattering matrix $S$ at zero excitation energy is constrained by both particle-hole symmetry and time-reversal symmetry. In the basis \eqref{eq:spgrading} (which relates time-reversed Majorana modes) the unitarity and symmetry constraints read\cite{Ful11}
\begin{equation}
S=S^{\ast}=-S^{T},\;\;S^2=-1,\label{symmetry}
\end{equation}
so the scattering matrix is real orthogonal and antisymmetric. (The superscript $T$ indicates the transpose.) 

The most general parameterization contains two real angles $\alpha,\vartheta\in (0,2\pi)$ and one $\mathbb{Z}_{2}$ index $\eta\in\{+1,-1\}$,\cite{note3}
\begin{align}
 &S = \begin{pmatrix}
      A\cos\alpha & -O^T\sin\alpha \\
      O\sin\alpha & -\eta\, A\cos\alpha
     \end{pmatrix},\label{general_s}\\
&O = \begin{pmatrix}
      -\cos \vartheta & -\eta \sin\vartheta \\
      \sin \vartheta & -\eta\cos\vartheta
     \end{pmatrix},\;\;
A=\begin{pmatrix}
0&1\\
-1&0
\end{pmatrix}.     
     \label{Odef}
\end{align}
The corresponding parameterization for $S'$ is obtained upon a permutation of the basis states,
\begin{equation}
S'=PSP^T,\;\;{\rm with}\;\; P= \begin{pmatrix}
     0 & 1 & 0 & 0 \\
     0 & 0 & 0 & 1 \\
     1 & 0 & 0 & 0 \\
     0 & 0 & 1 & 0
    \end{pmatrix},\label{SSprimerelation}
\end{equation}
which amounts to a $90^{\circ}$ rotation.

The angle $\alpha$ characterizes the scattering at a node for each of the two spin directions separately, while the angle $\vartheta$ couples them. For $\vartheta=0$ we have two independent, time-reversed, copies of the Cho-Fisher model,\cite{Cho97} representing a pair of uncoupled chiral superconductors of opposite chirality. The orthogonal matrix $O$ couples the two copies and produces a network model for a helical superconductor, in much the same way that Obuse et al.\cite{Obu07} obtained a network model for a quantum spin-Hall insulator by coupling a pair of Chalker-Coddington models\cite{Cha88} for quantum Hall insulators in opposite magnetic fields. 

\subsection{Vortices}
\label{vortices}

One difference between the superconducting network model considered here and the insulating model of Ref.\ \onlinecite{Obu07}, is that here the coupling of time-reversed networks is via real orthogonal rather than complex unitary matrices. This difference expresses the Majorana character of the modes, which have real rather than complex wave amplitudes. Another difference is the appearance of the $\mathbb{Z}_{2}$ index $\eta$, which determines the parity of the number of (time-reversally invariant) vortices in a plaquette. 

To see this, we take $\alpha=0$ or $\alpha=\pi$, when the network consists of isolated plaquettes. Denoting the value of $\eta$ for $S$ and $S'$ by $\eta_S$ and $\eta_{S'}$, the phase factor acquired by the Majorana mode as it encircles a plaquette is $\eta_{S}\eta_{S'}$. A bound state at zero excitation energy (doubly degenerate because of time-reversal symmetry) results if $\eta_{S}\eta_{S'}=1$. Since such Majorana zero-modes come in pairs, and each time-reversally invariant vortex in a helical superconductor traps one (doubly-degenerate) zero mode,\cite{Qi09} we conclude that the Majorana mode encircles an odd number of vortices for $\eta_{S}\eta_{S'}=1$.

In what follows we will assume that the system contains no vortices at all in the absence of disorder, so we choose $\eta_S=1$, $\eta_{S'}=-1$. The scattering matrices then take the form
\begin{subequations}
\label{SSprime}
\begin{align}
&S= \begin{pmatrix}
     0   & r   & t\cos\vartheta    & -t\sin\vartheta    \\
     -r  & 0   & t\sin\vartheta    & t\cos\vartheta     \\
     -t\cos\vartheta    & -t\sin\vartheta    & 0  & -r  \\
     t\sin\vartheta     & -t\cos\vartheta     & r  & 0
    \end{pmatrix},\label{SSprimea}\\
&S'=\begin{pmatrix}
     0  & -t\cos\vartheta    & -r  & -t\sin\vartheta    \\
    t \cos\vartheta     & 0   & t\sin\vartheta      & -r  \\
     r  & -t\sin\vartheta     & 0   & t\cos\vartheta    \\
    t \sin\vartheta   & r   & -t\cos\vartheta     & 0
    \end{pmatrix},\label{SSprimeb}
    \end{align}
\end{subequations}
where we have abbreviated $r=\cos\alpha$, $t=\sin\alpha$.

\subsection{Vortex disorder}
\label{disorder}

Disorder is introduced in the network model by varying the scattering parameters in a random way from one node to the other. We choose to keep the coupling strength $\vartheta$ the same for each node and to vary $\alpha$. Following the same procedure as for the Cho-Fisher model,\cite{Cha01} we draw $\alpha_{i}$ at each node $i$ independently from a distribution $P(\alpha_{i})$ given by
\begin{align}
 {\cal P}(\alpha_{i}) ={}& (1-q)\delta(\alpha_{i} - \alpha) +
\tfrac{1}{2}q\delta(\alpha_{i} + \alpha) \nonumber\\
 & + \tfrac{1}{2}q\delta(\alpha_{i}+\alpha-\pi).\label{eq:pdf}
\end{align}
The parameter $q \in [0,1]$ plays the role of disorder strength. 

This is a form of vortex disorder:\cite{Cha01} with probability $q$ two time-reversally invariant vortices are inserted in the plaquettes adjacent to the $i$-th node, one vortex in one plaquette and another one in the diagonally opposite plaquette. The diagonally opposite plaquettes are themselves chosen with equal probability $1/2$ from the two $\pm$ sublattices in Fig.\ \ref{fig:nodes}.

We use vortex disorder instead of purely electrostatic disorder [as in the Hamiltonians \eqref{HDdef} and \eqref{HDIIIdef}], because it scatters more effectively and allows us to localize wave functions in smaller systems. Both forms of disorder can produce Majorana zero-modes,\cite{Wim10,note4} so we do not expect qualitatively different features.

\section{Topological quantum number and thermal conductance}
\label{topquantnum}

A class DIII topological superconductor in two dimensions has a $\mathbb{Z}_{2}$ topological quantum number ${\cal Q}=\pm 1$. Formulas for ${\cal Q}$ exist based on the Hamiltonian\cite{Qi10} or on the scattering matrix.\cite{Ful12} Since the network model is described in terms of a scattering matrix, we use the latter formulation. 

We consider a rectangular geometry in the $x$-$y$ plane, of length $L={\cal N}'a$ in the $x$-direction and width $W={\cal N}a$ in the $y$-direction, where $a$ is the lattice constant and ${\cal N}$, ${\cal N}'$ are even integers (see Fig.\ \ref{fig:nodes}). In the transverse direction we impose either periodic boundary conditions, $\psi(x,0)=\psi(x,W)$, or antiperiodic boundary conditions, $\psi(x,0)=-\psi(x,W)$, on the wave functions. In the longitudinal direction we have absorbing boundary conditions,\cite{note5} corresponding to normal-metal reservoirs at $x=0$ and $x=L$.

The scattering matrix ${\cal S}$ of the entire structure has ${\cal N}\times {\cal N}$ reflection and transmission blocks,
\begin{equation}
{\cal S}=\begin{pmatrix}
{\cal R}&-{\cal T}^{T}\\
{\cal T}&{\cal R}'
\end{pmatrix}.\label{SRTdef}
\end{equation}
The reflection block is a real antisymmetric matrix. Its Pfaffian\cite{Wim12} determines the topological quantum number,\cite{Ful12}
\begin{equation}
{\cal Q} = {\rm sign}\,Q,\;\;Q= ({\rm Pf}\,{\cal R}_{\rm pbc})({\rm Pf}\,{\cal R}_{\rm apbc}),\label{eq:charge}
\end{equation}
where the labels pbc and apbc refer to the periodic and antiperiodic boundary conditions. (It does not matter if one takes ${\cal R}$ or ${\cal R}'$, they give the same ${\cal Q}$.)

The use of periodic (or antiperiodic) boundary conditions is convenient to minimize finite-size effects and is sufficient to study bulk properties. To study edge properties, one can impose reflecting boundary conditions by terminating the lattice as described in Ref.\ \onlinecite{Obu08}. Depending on how the lattice is terminated, one would then find that either ${\cal Q}=+1$ or ${\cal Q}=-1$ produces a helical edge state along the boundary, and therefore represents a topologically nontrivial phase. In the present study, without reflecting boundaries, we can distinguish different topological phases --- but we cannot decide which phase is trivial and which nontrivial.

In addition to the topological quantum number we calculate the two-terminal thermal conductance $G$ of the strip. This transport property is determined by the  transmission matrix ${\cal T}$,
\begin{equation}
G=G_{0}\,{\rm Tr}\,{\cal TT}^{\dagger},\label{Gthermaldef}
\end{equation}
with $G_{0}=\pi^{2}k_{B}^{2}T_{0}/6h$ the thermal conductance quantum and $T_{0}$ the temperature of the normal-metal reservoirs. The dimensionless thermal conductivity $g$ is defined by
\begin{equation}
g=(L/W)(G/G_{0}).\label{gdef}
\end{equation}
For the calculation of the thermal conductivity we take periodic boundary conditions in the $y$-direction and a large aspect ratio $W/L=4$, so that the choice of boundary conditions in the transverse direction has only a small effect.

\section{Topological phase transitions}
\label{topphasetrans}

\subsection{Phase diagram without disorder}
\label{nodisorder}

\begin{figure}[tb] 
\centerline{\includegraphics[width=1.0\linewidth]{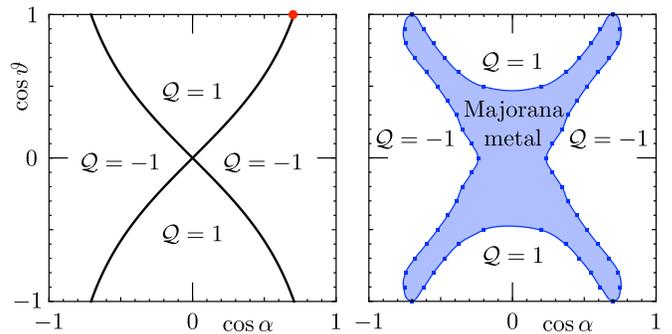}}
\caption{\label{figphaseclean} \textit{Left panel:} Phase diagram of the DIII network model without disorder. The solid curves, given by Eq.\ \eqref{criticalclean}, separate helical topological superconductors with different values ${\cal Q}$ of the $\mathbb{Z}_{2}$ topological quantum number. The red dot (at $\vartheta=0$, $\alpha=\pi/4$) marks the critical point in the Cho-Fisher model\cite{Cho97} of a chiral topological superconductor (class D). \textit{Right panel:} Phase diagram in the presence of disorder ($q=0.07$), when the topologically distinct phases are separated by a thermal (Majorana) metal.
}
\end{figure}

In the absence of disorder the location of the topological phase transitions can be determined exactly using a duality relation, see App.\ \ref{criticalpointderivation}. We find that the topological quantum number switches sign at critical points $\alpha_c,\vartheta_c$ that satisfy
\begin{equation}
|\sin\alpha_c\cos\vartheta_c|=|\cos\alpha_c|.\label{criticalclean}
\end{equation}
The phase boundaries are plotted in Fig.\ \ref{figphaseclean} (left panel), together with the values of the topological quantum number \eqref{eq:charge} at the two sides of the transition. For $\vartheta_c=0$ we recover the known value $\alpha_c = \pi/4$ of the critical point in the chiral Cho-Fisher model.\cite{Cho97} This is as expected, since for $\vartheta=0$ our helical network model consists of two independent chiral copies.

\subsection{Scaling of the critical conductivity}
\label{criticalcond}
 
\begin{figure}[tb]
\centerline{\includegraphics[width=0.9\linewidth]{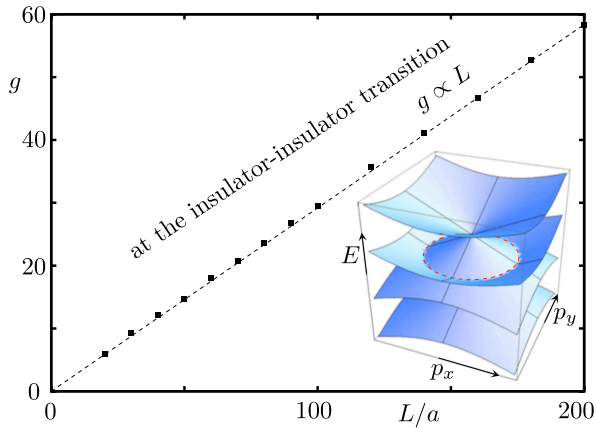}}
\caption{Data points (with dashed line as a guide to the eye): conductivity as a function of system size (at fixed aspect ratio $W/L=4$) of the DIII network model without disorder, at criticality for $\vartheta=0.4$. The ballistic scaling $g\propto L$ results from the Fermi circle of the long-wavelength Hamiltonian \eqref{eq:longH} (red dotted circle in the inset).
\label{fig:cg}}
\end{figure}

At the phase boundaries the excitation gap of the system closes. In the chiral network model this produces a scale-invariant thermal conductivity $g=1/\pi$, regardless of whether the system is disordered or not,\cite{Med10} but in our helical model the conductivity at the critical point scales ballistically $\propto L$ in the absence of disorder. The ballistic scaling is demonstrated in Fig.\ \ref{fig:cg} and can be understood by examining the long-wave length Hamiltonian corresponding to the network model.

The calculation of this Hamiltonian proceeds entirely along the lines of the analogous calculation for the quantum spin-Hall insulator,\cite{Ryu10a} so we only give the result. To first order in the deviation from the Cho-Fisher critical point $(\alpha,\vartheta)=(\pi/4,0)$, we find
\begin{align}
H={}&{\cal U}^{\dagger}\bigl[p_{x}\tau_{x}\sigma_{z}+p_{y}\tau_{y}\sigma_{0}+2(\alpha-\pi/4)\,\tau_{z}\sigma_{0}\nonumber\\
&+\sqrt{2}\,\vartheta\,\tau_{y}\sigma_{y}\bigr]{\cal U}\nonumber\\
={}&(p_x \tau_x + p_y \tau_y) \sigma_0 + 2(\alpha-\pi/4)\, \tau_z \sigma_z +\sqrt{2}\,\vartheta\, \tau_0\sigma_x,\label{eq:longH}\\
{\cal U}={}&\tfrac{1}{2}(\tau_{0}+i\tau_{y})\sigma_{0}+\tfrac{1}{2}(\tau_{0}-i\tau_{y})\sigma_{z}.\label{Udef}
\end{align}
Up to a unitary transformation ${\cal U}$, and to first order in (dimensionless) momentum $p$, this Hamiltonian has the form of $H_{\rm DIII}$ in Eq.\ \eqref{HDIIIdef}, with $\alpha$ playing the role of the chemical potential $\mu$ and $\vartheta$ playing the role of the spin coupling strength $K$.

The gap closes at $\alpha=\pi/4$ on a (twofold degenerate) circle $p_{x}^{2}+p_{y}^{2}=2\vartheta^2$ in momentum space (inset in Fig.\ \ref{fig:cg}). The Fermi wavevector $k_{F}=\sqrt{2}|\vartheta|/a$ corresponds to a ballistic conductance $G/G_{0}=2k_{F}W/\pi$. Hence we find the ballistic critical conductivity
\begin{equation}
g_{c}=\frac{2\sqrt{2}|\vartheta|L}{\pi a},\;\;|\vartheta|\ll 1.\label{gcdef}
\end{equation}

The ballistic scaling of the critical conductivity is a signature of the appearance of a Fermi circle at the phase transition, which is a special property of our model (chosen to maximize the coupling between the two opposite chiralities). More generally, the gap in a class-DIII Hamiltonian will close at four isolated points in momentum space,\cite{Ber10} resulting in a scale-invariant critical conductivity.

\subsection{Phase diagram with disorder}
\label{withdisorder}

\begin{figure}[tb]
\centerline{\includegraphics[width=0.8\linewidth]{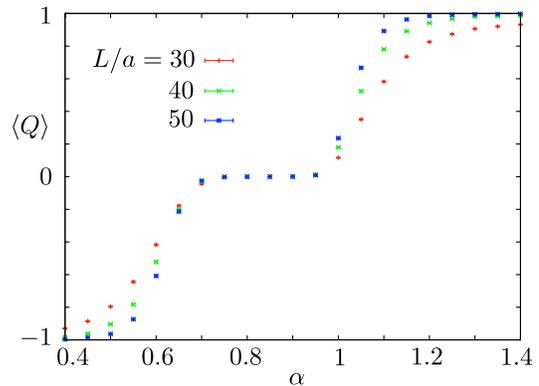}}
 \caption{Disorder average of $Q$, defined in Eq.\ \eqref{eq:charge} as a function of $\alpha$ for fixed $\vartheta=0.5$, $q=0.1$, and different values of $L=W/4$. The topological quantum number switches from $-1$ to $+1$ in the insulating phases via a plateau of zero average in the metallic phase.
\label{fig:charges}}
\end{figure}

\begin{figure}[tb]
\centerline{\includegraphics[width=0.8\linewidth]{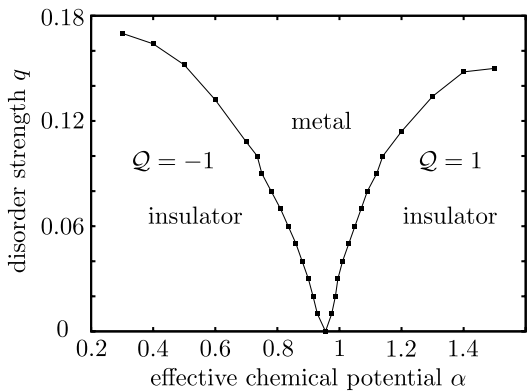}}
 \caption{Phase diagram of the disordered DIII network model for fixed $\vartheta = 0.785$. \label{fig:phasediagram}}
\end{figure}

As disorder is introduced in the system, a metallic phase develops in between the insulating phases, so that the switch from ${\cal Q}=+1$ to ${\cal Q}=-1$ goes via two metal-insulator transitions. In the metallic region ${\cal Q}$ has a random sign, averaging out to zero (see Fig.\ \ref{fig:charges}, where we averaged $Q$ rather than ${\cal Q}={\rm sign}\,Q$ to reduce statistical fluctuations). The phase diagram is shown in Fig.\ \ref{figphaseclean} (right panel) for a fixed disorder strength $q$ in the $\alpha$-$\vartheta$ plane and in Fig.\ \ref{fig:phasediagram} for fixed $\vartheta$ in the $\alpha$-$q$ plane. The metallic regions become broader and broader with increasing disorder, and for $q\gtrsim 0.2$ no insulating phase is left.

\begin{figure}[tb]
\centerline{\includegraphics[width=0.8\linewidth]{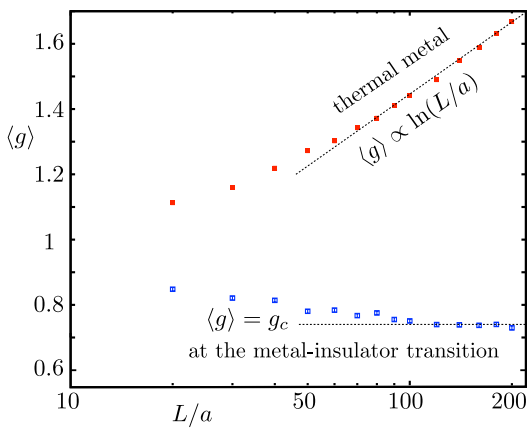}}
 \caption{Scaling of the conductivity in the disordered DIII network model ($\alpha=1.2$, $\vartheta=0.5$), at the metal-insulator transition (blue data points, $q=0.175$) and in the metallic phase (red data points, $q=0.2$). The dotted lines indicate the scale-invariance of the critical conductivity and the logarithmic scaling of the metallic conductivity.}
\label{fig_gscaling}
\end{figure}

Fig.\ \ref{fig_gscaling} shows the sample-size dependence of the average conductivity, both at the metal-insulator transition and in the metallic phase. (The exponential decay $\propto \exp(-L/\xi)$ in the insulating phase is not shown.) While without disorder the conductivity scales ballistically $\propto L$ at the critical point (see Fig.\ \ref{fig:cg}), disorder restores the scale invariance that is the hallmark of criticality. In the metallic phase we find a logarithmically increasing conductivity $\langle g\rangle=c \ln(L/a)$, characteristic for a Majorana metal,\cite{Sen00,Eve08,Med10}, with $c=1/\pi$ (dotted line in Fig.\ \ref{fig_gscaling}).

\subsection{Critical exponent}
\label{exponent}

The metal-insulator transition is associated with a diverging localization
length $\xi\propto|x-x_{c}|^{-\nu}$, where $x$ can be any of the control
parameters $\alpha,\vartheta,q$ and $x_{c}$ is the value of $x$ at the critical
point. To determine the critical exponent $\nu$ we perform a finite-size
scaling analysis of the thermal conductivity, in a manner
analogous to the work of Slevin and Ohtsuki.\cite{Sle99}

\begin{figure}[tb]
\centerline{\includegraphics[width=0.8\linewidth]{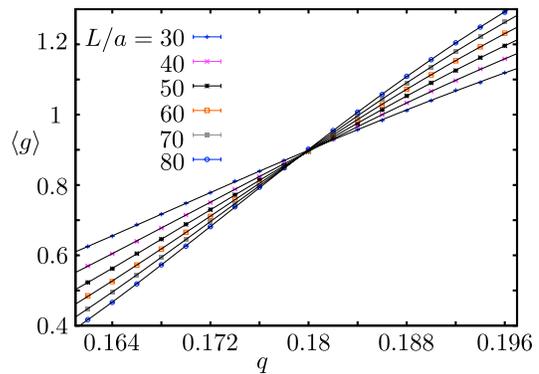}}
\caption{Disorder-averaged conductivity as a function of disorder strength for various system sizes $L=W/4$ at fixed $\alpha = 1.2$, $\vartheta=0.5$. Solid curves are a fit to the scaling law, as described in App.\ \ref{finitesize}. The curves cross at the critical point.\label{fig:t04d04}}
\end{figure}

Typical data is shown in Fig.\ \ref{fig:t04d04}, where we follow the thermal conductivity through the metal-insulator transition upon varying the disorder strength $q$ at fixed $\alpha$ and $\vartheta$. The dimensions $L$ and $W$ of the system are increased at fixed aspect ratio $W/L=4$. The curves are fits of the data to the scaling law, as described in App.\ \ref{finitesize}. Results are given in Table \ref{tablescaling}.

\begin{table}[htb]
\centering
\begin{tabular}{ | c | c | c | c | c | c | c | c | c |}
\hline
control parameter& $\nu$ & $g_c$ \\ \hline\hline
$x\equiv q$, $\alpha=1.2$, $\vartheta=0.5$ & $2.06$ $[1.89, 2.20]$ &  $0.74$ $[0.66, 0.81]$ \\ \hline
$x\equiv \alpha$, $q=0.1$, $\vartheta=0.785$ & $1.93$ $[1.78, 2.24]$ & $0.73$ $[0.69, 0.76]$ \\ \hline
\end{tabular}
\caption{Critical exponent $\nu$ and critical conductivity $g_{c}$, with $\pm 3\sigma$ confidence levels indicated by square brackets, extracted from a finite-size scaling analysis with $x=q$ or $x=\alpha$ as control parameter to drive the system through a metal-insulator transition. The two cases correspond to transitions into a topological phase with opposite value of ${\cal Q}$.}
\label{tablescaling}
\end{table}

\section{Conclusion}
\label{conclude}

\begin{table*}[htb]
\centering
\begin{tabular}{ | c | c | c | c | c | c | c | c | c |}
\hline
& symmetry & time-rev.\ & spin-rot.\ & topological & insul.-insul.\ & metal-insul.\ & Refs.\ \\ 
& class  & symmetry & symmetry & quantum nr.\ & transition & transition &  \\ \hline\hline
quantum Hall insulator & A & $\times$ & $\times$ & $\mathbb{Z}$ & $\nu\approx
2.6$ & --- & \onlinecite{Cha88,Sle09} \\ \hline
quantum spin-Hall insul.\ & AII & $\checkmark$ & $\times$ & $\mathbb{Z}_{2}$ & --- & $\nu\approx 2.7$ & \onlinecite{Asa04,Mar06,Obu07,Kob11,Nie12} \\ \hline
chiral \textit{d}-wave supercond.\ & C & $\times$ & $\checkmark$ & $\mathbb{Z}$ & $\nu=4/3$ & --- &  \onlinecite{Gru99,Bea02} \\ \hline
chiral \textit{p}-wave supercond.\ & D & $\times$ & $\times$ & $\mathbb{Z}$ & $\nu=1$ & $\nu\approx 1.0$ & \onlinecite{Med10,Lud94} \\ \hline
helical \textit{p}-wave supercond.\ & DIII & $\checkmark$ & $\times$ & $\mathbb{Z}_{2}$ & --- & $\nu\approx 2.0$ & \textit{this work} \\ \hline
\end{tabular}
\caption{Overview of the critical exponents in the five symmetry classes that exhibit a topological phase transition in two dimensions.}
\label{tablenu}
\end{table*}

In conclusion, we have presented a network model for a two-dimensional helical \textit{p}-wave superconductor and used it to investigate the topological phase transitions. The presence of time-reversal symmetry (class DIII) leads to differences with the more familiar chiral \textit{p}-wave superconductors (class D), of a qualitative nature (the appearance of a thermal metal separating the thermally insulating phases) and also of a quantitative nature (an approximate doubling of the critical exponent). Helical superconductors have not yet been convincingly demonstrated in experiments, but there is an active search and a variety of candidate materials.\cite{Tan09,Sat09,Liu11,Won11,Nak12}

This study fills in the last missing entry in the list of critical exponents of two-dimensional topological phase transitions (see Table \ref{tablenu}), completing a line of  research on network models that started with the seminal work of Chalker and Coddington on the quantum Hall effect.\cite{Cha88} It is intriguing that the effect of time-reversal symmetry is close to a doubling of the critical exponent (from $\nu\approx 1.0$ in class D to $\nu\approx 2.0$ in class DIII), but this may well be accidental.

\acknowledgments
We received help with the numerics from C. W. Groth, K. Slevin, and M. Wimmer.
This research was supported by the Dutch Science Foundation NWO/FOM, by EPSRC Grant EP/F032773/1, by an ERC Advanced Investigator grant, and by the EU network NanoCTM.

\appendix

\section{Location of the critical point in the network model without disorder}
\label{criticalpointderivation}

The critical point in the clean DIII network model ($q=0$) can be obtained from a duality relation: Exchange of the scattering matrices $S$ and $S'$ of the two sublattices (black and white nodes in Fig.\ \ref{fig:nodes}) has the effect of exchanging the trivial and nontrivial phases. This can be seen most easily for reflecting boundary conditions, when exchange of the sublattices either creates or removes the helical edge state (see Fig.\ 7 of Ref.\ \onlinecite{Obu08}). 

The exchange of $S$ and $S'$ amounts to the transformation of $\alpha,\vartheta$ into $\alpha',\vartheta'$, given by
\begin{align}
&\cos\alpha'=-\sin\alpha\cos\vartheta,\;\;\cos\alpha=-\sin\alpha'\cos\vartheta',\nonumber\\
&\sin\alpha'\sin\vartheta'=\sin\alpha\sin\vartheta.\label{alphaalphaprime}
\end{align}
Equivalently, the unit vector $\hat{n}=(\sin\alpha\cos\vartheta,\sin\alpha\sin\vartheta,\cos\alpha)$ is transformed into
\begin{equation}
n'_x=-n_z,\;\;n'_y=n_y,\;\;n'_z=-n_x,\label{nnprime}
\end{equation}
which amounts to a reflection in the plane $x+z=0$. The network is selfdual if $\hat{n}$ lies in this plane. Since a selfdual network is at the critical point, we arrive at a sufficient condition for criticality, $n_x+n_z=0$, or equivalently
\begin{equation}
\sin\alpha_c\cos\vartheta_c+\cos\alpha_c=0.\label{criticalclean1}
\end{equation}

An alternative condition can be obtained by noting that the transformation $\alpha\mapsto-\alpha$ leaves the reflection matrix ${\cal R}$ unaffected. The topological quantum number \eqref{eq:charge} therefore remains unchanged, so if $\alpha_c,\vartheta_c$ is critical then also $-\alpha_c,\vartheta_c$. We thus have a second sufficient condition for criticality,
\begin{equation}
-\sin\alpha_c\cos\vartheta_c+\cos\alpha_c=0.\label{criticalclean2}
\end{equation}
Eqs.\ \eqref{criticalclean1} and \eqref{criticalclean2} together give the condition \eqref{criticalclean}.

\section{Finite-size scaling analysis}
\label{finitesize}

\begin{table*}[htb]
\centering
\begin{tabular}{ | c | c | c | c | c | c |}
\hline
control & fit parameters & irrelevant exp.\ & reduced & nr.\ of degrees& goodness \\ 
parameter & & & $\chi^2$ & of freedom & of fit \\ \hline\hline
$x\equiv q$& $n=m_1=1$, $m_0=3$ & $y = -0.44$ & $1.04$ & $97$ & $0.65$ \\ 
$\alpha=1.2$, $\vartheta=0.5$& $q_r=2$, $q_i=0$ &  $[-0.81, -0.30]$ & & &\\ \hline
$x\equiv \alpha$ & $n=1$, $m_0 = m_1=2$ & $y = -0.67$& $1.06$ & $128$ & $0.56$ \\
$q=0.1$, $\vartheta=0.785$& $q_r=3$, $q_i=1$ &  $ [-0.78, -0.53]$ & & & \\ \hline
\end{tabular}
\caption{Parameters for the nonlinear fitting analysis, giving the critical exponent and conductivity of Table \ref{tablenu}.}
\label{tablefit}
\end{table*}

We determine the critical exponent $\nu$ associated with the metal-insulator transitions on both sides of the metallic phase, by an analysis of the system size $L=W/4$ dependence of the disorder averaged conductivity $g$. Following the general approach of Slevin and Ohtsuki,\cite{Sle99} we take into account finite-size corrections to scaling in the form of nonlinearities in the scaling variable $u$, as well as the presence of an irrelevant scaling exponent $y<0$.

The finite-size scaling law reads
\begin{equation}
 g = F(u_0 L^{1/\nu}, u_1 L^y),
\end{equation}
in terms of the relevant scaling variable $u_0$ and the leading irrelevant scaling variable $u_1$. We perform a Taylor expansion, first up to order $n$ in powers of $u_1$,
\begin{equation}
 g= \sum_{k=0}^n u_1^kL^{ky}F_k(u_0 L^{1/\nu}),
\end{equation}
and then on each of the functions $F_k$ up to order $m_k$ in powers of $u_0$,
\begin{equation}
 F_k(u_0 L^{1/\nu}) = \sum_{j=0}^{m_k} u_0^jL^{j/\nu}F_{kj}.
\end{equation}

We tune through the metal-insulator transition by varying one parameter $x\in\{\alpha,\vartheta,q\}$ through the critical point $x_c$, keeping the other two parameters fixed. Nonlinearities are taken into account by Taylor expanding the relevant and irrelevant scaling variables in powers of $x-x_c$, up to orders $q_r$ and $q_i$ respectively,
\begin{align}
& u_0( x -  x_c) = \sum_{k=1}^{q_r} b_k ( x -  x_c)^k,\\
& u_1( x -  x_c) = \sum_{k=0}^{q_i} c_k ( x -  x_c)^k.
\end{align}
The expansion of the relevant scaling variable does not contain a zeroth order term, due to the requirement $u_0(0)=0$ for a scale-invariant critical conductivity.

The average conductivity is determined up to a precision between $\sim 0.2\%$ and $\sim 0.07\%$ (error bars much smaller than the size of the symbols in Fig. \ref{fig:t04d04}). We perform the fit by minimizing the $\chi^2$ statistic, and express the goodness of fit as well as the degree of uncertainty in the fit parameters through a Monte Carlo resampling technique,\cite{Pre82} as appropriate for a non-linear fitting function. Results are collected in Tables \ref{tablescaling} and \ref{tablefit}.


\begin{thebibliography}{99}
\bibitem{note3} The parameterization \eqref{general_s} follows from the general polar decomposition of a class DIII scattering matrix of dimension $2{\cal N}\times 2{\cal N}$ given in Ref.\ \onlinecite{Ful11}. The ${\cal N}=2$ result is simplified by means of the identity $OAO^{T}=\eta A$, with $\eta={\rm Det}\,O$.
\bibitem{note5} The orientation of the network in Fig.\ \ref{fig:nodes} is chosen such that the interface resistance at the absorbing boundaries $x=0,L$ is minimized. A $45^{\circ}$ rotation is more convenient to create reflecting boundaries at $y=0,W$, but gives a substantially larger interface resistance at the absorbing boundaries.
\end{thebibliography}
\end{document}